\documentclass[useAMS,usenatbib]{mn2e}
\usepackage{graphicx}
\begin{document}

\title[Characterisation of AstraLux binaries]{Characterisation of close visual binaries from the AstraLux Large M Dwarf Survey\thanks{Based on observations made with ESO telescopes at the Paranal Observatory under programme IDs 
086.A-9014,
086.C-0869, 
088.A-9032,
088.C-0753,
089.A-9013 and
091.C-0216.
}}
\author[C. Bergfors et al.]
{C. Bergfors$^{1,2}$\thanks{E-mail:c.bergfors@ucl.ac.uk}, W. Brandner$^{2}$, M. Bonnefoy$^{3,4}$, J. Schlieder$^{5}$, M. Janson$^{6}$, 
\newauthor Th. Henning$^{2}$ and G. Chauvin$^{3,4}$\\
$^{1}$Department of Physics and Astronomy, University College London, 132 Hampstead Road, London NW1 2PS, United Kingdom\\
$^{2}$Max-Planck-Institut f\"ur Astronomie, K\"onigstuhl 17, 69117 Heidelberg, Germany \\
$^{3}$Univ. Grenoble Alpes, IPAG, F-38000 Grenoble, France\\
$^{4}$CNRS, IPAG, F-38000 Grenoble, France\\
$^{5}$NASA Postdoctoral Program Fellow, NASA Ames Research Center, MS 245-6, Moffett Field, CA 94035, USA\\
$^{6}$Institutionen f\"or astronomi, Stockholms universitet,106 91 Stockholm, Sweden}

\date{Accepted 2015 November 23. Received 2015 October 9; in original form 2015 May 15}

\pagerange{\pageref{firstpage}--\pageref{lastpage}} \pubyear{2015}

\maketitle

\label{firstpage}

\begin{abstract}
We present VLT/SINFONI $J, H+K$ spectra of seven close visual pairs in M dwarf
binary/triple systems, discovered or observed by the AstraLux M dwarf survey.
We determine the spectral types to within $\pm1.0$ subclasses from comparison
to template spectra and the strength of $K$-band water
absorption, and derive effective temperatures. The results are compared to optical spectral types of the unresolved binary/multiple systems, and we confirm that our photometric
method to derive spectral types in the AstraLux M dwarf survey is accurate.
We look for signs of youth such as chromospheric activity and low surface gravity, and find
an age in the range $0.25-1$\,Gyr for the GJ\,852 system.
Strong Li absorption is detected in optical spectra of the triple system J024902 obtained with FEROS at the ESO-MPG 2.2\,m telescope. The equivalent width of the absorption suggests an age consistent with the $\beta$ Pic moving group. However, further observations are needed to establish group membership.
Ongoing orbital monitoring will provide dynamical masses and thus calibration of
evolutionary models for low mass stars.

\end{abstract}


\section{Introduction}

M dwarfs in multiple systems can provide valuable insight into the structure, formation and evolution of very-low mass stars and brown dwarfs, through their multiplicity characteristics as well as their physical and orbital properties \citep[e.g.,][]{Burgasser2007,Goodwin2007, Janson2007, DucheneKraus2013}. Despite being the most common stars in our neighbourhood, fundamental physical characteristics 
such as mass, radius, luminosity, and relations between these properties,
are not as well constrained for mid- to late-M type dwarfs as for solar-type and intermediate mass stars. 
Detailed studies of orbital elements of individual low mass binaries, and dynamical masses, are needed for empirical calibration of models for low-mass stars, and significant effort has therefore been made in recent years to better characterise in particular ultra-cool dwarfs \citep[e.g.][] {Bouy2008, Bonnefoy2009, Konopacky2010, Dupuy2010, Schlieder2014, Zhou2014}.
In addition, the recent discoveries and dedicated surveys for M dwarf planets require accurate stellar physical properties in order to derive reliable planetary parameters, furthering the interest for characterising M dwarfs \citep[e.g.,][]{Johnson2012, Mann2012, Mann2013b, Mann2013,  Muirhead2012, Muirhead2014, DressingCharbonneau2013, Newton2014, Newton2015, Gaidos2014, Alonso-Floriano2015, Bowler2015}.

Atmospheric and evolutionary models are particularly poorly constrained for young 
($\la100$\,Myr) M dwarfs, of which only a handful of binaries have measured dynamical masses 
\citep[see, e.g.,][]{Close2005, Bonnefoy2009}. 
In order to better constrain multiplicity properties and identify young binaries suitable
 for dynamical mass measurements, we carried out the AstraLux large
  M dwarf survey: a Lucky Imaging multiplicity survey of 761 young, nearby 
  late-K and M dwarfs \citep{Bergfors2010, Janson2012}, supplemented by 286 mid- to 
  late-M dwarfs in an extension of the survey \citep{Janson2014a}. 
   From this survey we selected seven pairs in binary or triple systems for the spectroscopic observations presented in this paper, to better characterise the stars with respect to spectral types and youth. A large fraction of the AstraLux targets, including five of the binaries 
studied here, have recently been kinematically linked to young associations 
\citep[][Schlieder et al., in preparation]{Malo2013, Malo2014a, Rodriguez2013}.
Observed properties such as spectral type, surface gravity and 
luminosity can be compared to evolutionary models of pre-main sequence stars to yield component mass and system age 
estimates for close binaries. Ongoing astrometric and radial velocity monitoring will 
provide a better understanding of these objects from well determined orbits and dynamical 
masses within a few years, which will then be compared to the results presented here.

This paper is organised as follows: After a short introduction in Section 1, Section 2 describes the target selection, observations and data reduction procedure. In Section 3, the individual component spectral types and effective temperatures are determined and compared to previous estimates. We also measure the equivalent widths of gravity sensitive features and compare to old field stars and to stars in young associations. 
We discuss our results and additional youth indicators in Section 4, and compare the observed and derived $M_K$ vs. $T_{\rm eff}$ of GJ\,852\,BC and J061610 to theoretical isochrones to infer individual masses and system ages. We end this report with a summary of our results and conclusions in Section 5.

\section{Observations and data reduction}
\subsection{Target selection and observations}
Seven nearby M dwarf binaries or close pairs in hierarchical triple systems were selected from the target list of the AstraLux M 
dwarf survey as good candidates for follow-up near-infrared spectroscopy and characterisation. These are listed in Table  \ref{observations}.
The selected targets had all been observed in at least two epochs and were confirmed as physically bound via common proper motion. They all had a projected separation of 
$a<8$\,AU and primaries with photometric spectral types M3.5 or later derived from AstraLux observations in SDSS $i^\prime-$ and 
$z^\prime-$band. 
The stars in the AstraLux survey are suspected to be young, based on their coronal activity and low tangential velocity, and have spectroscopic distances of less than 52\,pc from the Sun \citep{Riaz2006}. The GJ\,852\,BC system has a USNO parallax distance of only 10\,pc \citep{HarringtonDahn1980}. 

 The targets were observed in service mode with the adaptive optics fed Spectrograph 
for INtegral Field Observations in the Near Infrared \citep[SINFONI,][]{Bonnet2004} at the VLT Unit Telescope 4 (Yepun). 
SINFONI consists of the SPIFFI integral field spectrograph \citep{Eisenhauer2003} together with the 
Multi-Application Curvature Adaptive Optics module \citep[MACAO,][]{Bonnet2003}.
We used the $J$ ($\lambda=1.1-1.4\,\mu$m) and $H+K$ ($\lambda=1.45-2.45\,\mu$m) gratings, with a resolving power 
of R$\approx$2000 and 1500 respectively, and the target itself as a natural guide star. 
Pre-slit optics were selected depending on binary separation to provide a spaxel scale of 
$\rm12.5\,mas\times25\,mas$, corresponding to a Field of View (FoV) of $0.8$ arcsec$^2$, for all the targets with angular separations $0.\arcsec1-0.\arcsec2$ (see Table \ref{observations}). For the wider couple, GJ\,852\,BC, the
$\rm125\,mas\times250\,mas$ spaxel pre-optics were used, corresponding to a FoV of 8 arcsec$^2$. 
Each target was observed in a dither sequence with small offsets between eight science exposures, and three sky dither points at the beginning, middle and end of science acquisition.
For each observation a telluric standard star of spectral type B or early A was observed at similar airmass. All observations were performed at airmass close to 1.0.
Table \ref{observations} lists the observational details: which components were observed (GJ\,852\,BC, J053018\,AB and J024902\,BC are part of 
triple systems with a wider companion), their angular separation as measured by \citet{Bergfors2010, Janson2012}, the date of observation, integration times and number of integrations, measured Signal-to-Noise ratio (S/N, see Sect. 3.1) and the spectral type of the telluric standard.

\begin{table*}
\caption{SINFONI observational details.}
\centering
\begin{tabular}{l r r r r r r r r l}
\hline\hline
2MASS ID	&	Comp.\footnotemark[1] & $\rho$\footnotemark[1]	&	Obs. Date	 &	Exp. Time	 &	Exp. Time	& $J$-band S/N	&	$H$-band S/N	& $K$-band S/N 		&Telluric Std	\\
		&						     &	[\arcsec]			&			&		$J$	 &	$H+K$		&	Prim/Sec	&	Prim/Sec	&	Prim/Sec		&		SpT	\\
\hline
J02133021-4654505	&	A, B	&	0.138	&	2011-11-15	&		$72\times20$\,s	&	$72\times6$\,s	&	51/26	&	58/60	&	90/82	&	B9\,V \\
J02490228-1029220	&	B, C	&	0.157	&	2010-12-01	&		$72\times5$\,s		&	$72\times2$\,s	&	20/20	&	34/34	&	39/33	&	B8\,V	\\
J04080543-2731349	&	A, B	&	0.221	&	2011-01-09	&		$72\times5$\,s		&	$72\times2$\,s	&	18/14	&	17/22	&	23/15	&	B3\,V	\\
J05301858-5358483	&	A, B	&	0.232	&	2011-11-16	&		$72\times6$\,s		&	$72\times2$\,s	&	92/90	&	73/61	&	95/69	&	B8\,V	\\	
J06134539-2352077	&	A, B	&	0.145	&	2011-11-16	&		$72\times7$\,s		&	$72\times2$\,s	&	97/64	&	67/67	&	74/58	&	B8\,V	\\
J06161032-1320422	&	A, B	&	0.194	&	2010-12-01	&		$72\times15$\,s	&	$72\times2$\,s	&	43/22	&	26/15	&	22/9		&	B9\,V	\\
J22171899-0848122	&	B, C	&	0.970	&	2010-10-21	&		$72\times2$\,s		&	$72\times2$\,s	&	63/32	&	86/56	&	95/45	&	A0\,V	\\

\hline
\label{observations}
\end{tabular}
\newline
\footnotetext{1}{$^1$ See \citet{Bergfors2010, Janson2012} for component definitions and astrometric measurements.}\\

\end{table*}

\subsection{Data reduction}
$J$ and $H+K$ band datacubes were built from a set of raw data and associated calibration frames using the  SINFONI data reduction pipeline version 2.2.5 \citep{Abuter2006}.
Some of the raw frames were affected by stripes on slitlet $\#25$ 
(the so called \textit{odd/even effect}\footnote[1]{http://www.eso.org/observing/dfo/quality/SINFONI/qc/dark\_QC1.html}) 
and by dark horizontal 
lines. These electronic artefacts were properly removed using custom scripts before providing the frames to 
the pipeline. The data reduction was checked by eye and 
a set of quality control parameters were compared to reference 
values\footnote[2]{http://www.eso.org/observing/dfo/quality/SINFONI/qc/qc1.html}. 
 
Though the binaries are successfully resolved by MACAO, cross contamination of their spectra is an issue that must be addressed.
We used a custom spectral extraction tool to deblend the flux of the components, and consequently their spectra, slice by slice in each of the datacubes. 
The tool is a modified version of the algorithm presented in \citet{Bonnefoy2009}. It first estimates the positions of the sources inside the 
FoV, which usually drift due to the atmospheric refraction\footnote[3]{The refraction can be corrected by the data reduction pipeline. However, we still noticed some residual drift after this correction had been applied.}. We then applied on each slice a modified version of the \citet{Dumas2001} CLEAN algorithm to retrieve the 
individual flux of the sources. The algorithm requires the PSF at the time of the observation and for each cube wavelength. 
To provide that, we 
considered two different approaches. We first used a scaled version of the telluric standard star data cubes observed immediately after our targets 
(hereafter \textit{PSFstd}). Alternatively, we built the PSF by duplicating the profile (hereafter \textit{PSFdup}) of the brightest binary component 
(or of the component farthest from the FoV edges). We re-estimated the position of the sources, and re-built the PSF in case 
\textit{PSFdup} was chosen, applying a second layer of CLEAN. 
For each input cube, an extracted cube for each binary component, and a residual map that enables to monitor the efficiency of the extraction process 
was produced. \textit{PSFdup} provides a more accurate extraction. On the contrary, the PSF shape built following \textit{PSFstd} is less appropriate, but the resulting spectra have slightly higher S/N in some cases. Since our spectral analysis is based mainly on the continuum shape (see Section 3.1), we used the \textit{PSFdup} reduced spectra for our analysis.

The science target and telluric standard spectra were extracted 
using the same aperture sizes for target and standard star so as to minimise differential flux losses. Aperture sizes of 225\,mas were considered optimal for obtaining high S/N for the small plate scale observations, and a 900\,mas aperture was used for GJ\,852\,BC.  
For each set of observations,  
the strong Br-series absorption lines in $H+K$ and the Pa$\beta$ line in $J$-band in our late-B -- early-A  telluric standard spectra were fitted with Voigt functions and subtracted from the standard star spectrum before it was divided 
by a blackbody function of corresponding temperature.
The science spectrum was then divided by the resulting
spectral response, containing only remaining instrumental and atmospheric features, to obtain the final spectrum.

\section{Results}
\subsection{Spectral type and effective temperature}

The strong atomic and molecular features and temperature sensitive continuum regions make $K$-band the optimal choice for determining spectral types of cool stars in the near-infrared.
As a first estimate, we performed a visual comparison of the $K$-band spectral shapes to template SpeX spectra obtained from the IRTF Spectral Library \citep{Cushing2005, Rayner2009}. We also compared
the strengths of some of the most prominent atomic and molecular absorption features in $H$- and $K$-band suitable for spectroscopic classification, such as Na, Ca, Al, Mg, Si, Fe, and CO and FeH band heads \citep{Cushing2005}, to the IRTF templates.

The S/N was measured in each wavelength band in small regions without prominent spectral features as suggested by \citet{Covey2010}, and are listed in Table \ref{observations}. Note that these values are representative and vary over the spectral range. The low S/N of the J040805 and J061610 spectra makes it difficult to assign precise spectral types to these targets based on absorption feature strengths, and we estimated rough spectral types for these stars from the continuum shapes in the $K$-band. 

\begin{figure}
 \centering
 \includegraphics[width=3.5in]{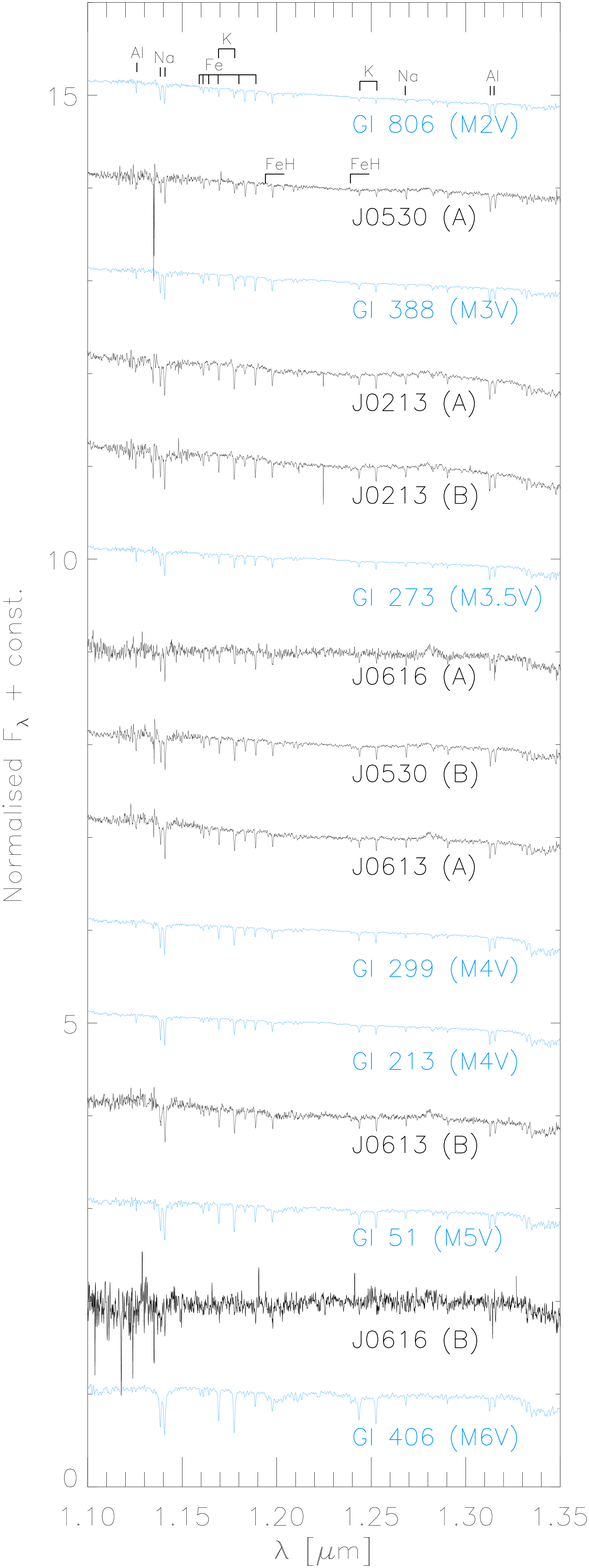} 
    \caption{$J$-band spectra. The SINFONI targets are shown in black, and IRTF templates in blue.}
\label{Jbandspectra}
 \end{figure}

Figures 1--4 show our $J$ and $H+K$ band spectra plotted together with the SpeX templates. 
For this and following analysis, we used the spectra extracted using the \textit{PSFdup}-method since it best preserves the individual spectral shapes. The $J$-band spectra of J024902 and J040805 are of low S/N ($\la20$) and are omitted from Figure  \ref{Jbandspectra} as no atomic features could be confidently identified.

The initial estimate is complemented by a more quantitative spectral type analysis in which we calculated the $\rm H_2O-K2$-index defined by \citet{Rojas-Ayala2012}:
\begin{equation}
H_2O-K2=\frac {\langle F(2.070-2.090)\rangle/\langle F(2.235-2.255)\rangle}{\langle F(2.235-2.255)\rangle/\langle F(2.360-2.380)\rangle}
\end{equation}

The index is a modified version of the \citet{Covey2010} $\rm H_2O-K$-index which takes into account the Mg \begin{small}I\end{small} and Ti \begin{small}I\end{small} atomic features that affect the measurements in bright spectra. It measures the temperature dependent strength of $K$-band water absorption and is independent of gravity and metallicity for $3000\,K\le T_{\rm eff}\le 3800$\,K. We measured the median flux in each wavelength region in Eq. 1 and estimated errors with a Monte Carlo simulation. Random Gaussian noise based on the S/N was added to a set of 10\,000 median flux measurements, from which the mean $\rm H_2O-K2$ index was calculated with error bars corresponding to the standard deviation of the resulting distribution. The indices and corresponding spectral types are listed in Table \ref{SpT}, together with the spectral types derived from $i^\prime-z^\prime$ by \citet{Bergfors2010, Janson2012} and the visually estimated $K$-band spectral types.

We derive the effective temperatures using the spectral type -- effective temperature relations for $600$\,Myr old M dwarfs from \citet{KrausHillenbrand2007}. The results are listed in Table \ref{SpT} and assume errors in $T_{\rm eff}$ directly corresponding to the estimated errors of the adopted spectral type. 

\begin{figure}
 \centering
 \includegraphics[width=3.5in]{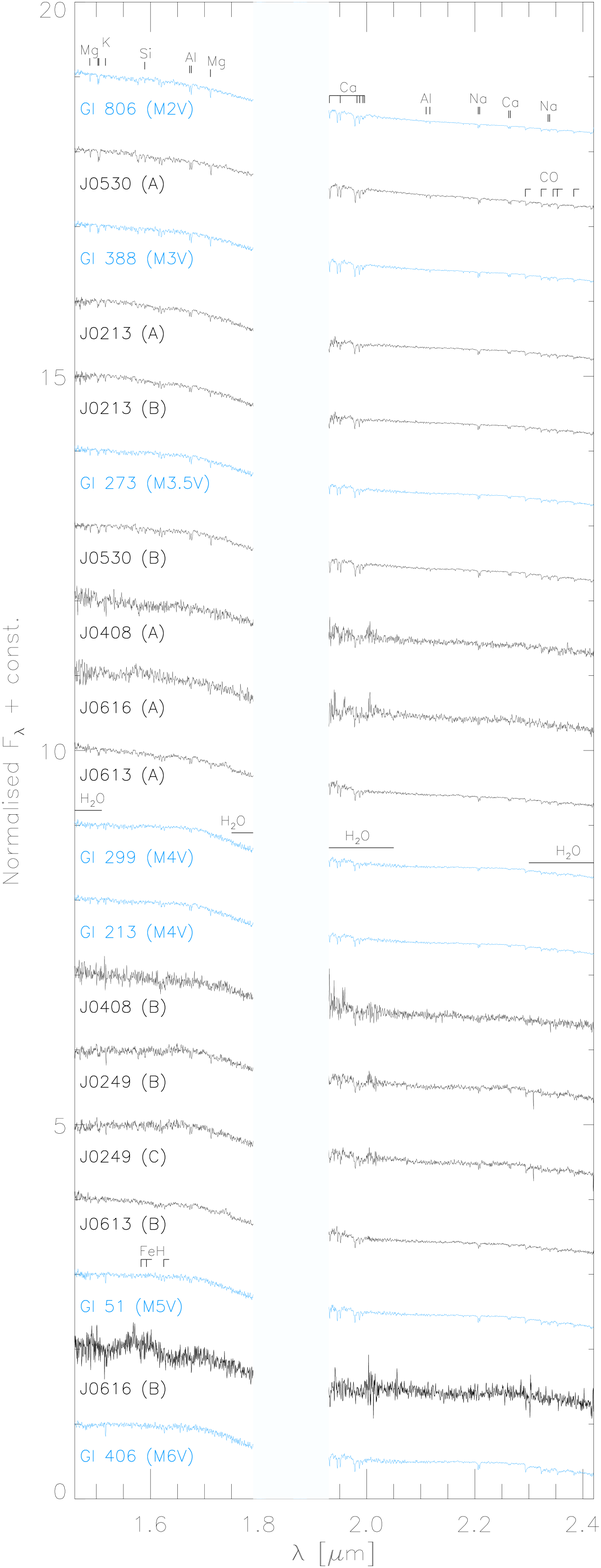} 

    \caption{Similar to Fig. \ref{Jbandspectra} but for $H+K$-band.}
\label{HKbandspectra}
 \end{figure}

\begin{table*}
\caption{Photometric and near-infrared spectral types, and inferred effective temperatures.}
\centering
\begin{tabular}{l  r r r l r l}
\hline\hline
ID	&	($i'-z'$) SpT		&	NIR SpT	 	& 	H$_2$O-K2	&	NIR SpT 		&	Adopted SpT  & $\rm T_{eff}$\\
	&	Photometry		&	K-band comp.	&				&	H$_2$O-K2	&			      & (K)		\\
\hline

J021330 (A)	&	M4.0 ($3.0-5.0$)	&	M3.0$-$M3.5	&	$0.89\pm0.03$	&	M3.5 ($3.0-4.0$)	&	M3.5 ($3.0-4.0$)	&	$3265\pm85$	\\
J021330 (B)	&	M5.0 ($4.0-6.0$)	&	M3.0$-$M3.5	&	$0.89\pm0.03$	&	M3.5 ($3.0-4.0$)	&	M3.5 ($3.0-4.0$)	&	$3265\pm85$	\\
J024902 (B)	&	M3.5 ($2.5-4.5$)	&	M4.0			&	$0.86\pm0.08$	&	M4.0 ($2.5-5.5$)	&	M4.0 ($3.0-5.0$)	&	$3180\pm170$	\\
J024902 (C)	&	M3.5 ($2.5-4.5$)	&	M4.0$-$M4.5	&	$0.87\pm0.08$	&	M4.0 ($2.0-6.0$)	&	M4.0 ($3.0-5.0$)	&	$3180\pm170$	\\	
J040805 (A)	&	M3.5 ($2.5-4.5$)	&	M3.0$-$M3.5	&	$0.99\pm0.14$	&	M1.0 (-$2.0-4.5$)	&	M3.5 ($2.5-4.5$)	&	$3265^{+165}_{-170}$	\\
J040805 (B)	&	M4.5 ($3.5-5.5$)	&	M3.0$-$M4.0	&	$0.90\pm0.14$	&	M3.5 ($0.0-6.5$)	&	M4.0 ($3.0-5.0$)	&	$3180\pm170$	\\
J053018 (A)	&	M3.0 ($2.0-4.0$)	&	M2.0$-$M3.0	&	$0.98\pm0.03$	&	M1.5 ($0.5-2.0$)	&	M2.0 ($1.0-3.0$)	&	$3510^{+170}_{-160}$	\\
J053018 (B)	&	M4.0 ($3.0-5.0$)	&	M3.0			&	$0.89\pm0.03$	&	M3.5 ($3.0-4.0$)	&	M3.5 ($3.0-4.0$)	&	$3265\pm85$	\\
J061345 (A)	&	M3.5 ($2.5-4.5$)	&	M3.5			&	$0.90\pm0.03$	&	M3.5 ($2.5-4.0$)	&	M3.5 ($2.5-4.0$)	&	$3265^{+165}_{-85}$	\\
J061345 (B)	&	M5.0 ($4.0-6.0$)	&	M3.5$-$M4.5	&	$0.86\pm0.04$	&	M4.0 ($3.5-5.0$)	&	M4.0 ($3.5-5.0$)	&	$3180^{+85}_{-170}$	\\
J061610 (A)	&	M3.5 ($2.5-4.5$)	&	M3.0$-$M4.0	&	$0.87\pm0.14$	&	M4.0 ($0.5-7.5$)	&	M3.5 ($3.0-4.5$)	&	$3265^{+85}_{-170}$	\\
J061610 (B)	&	M5.0 ($4.0-6.0$)	&	M4.0$-$M6.0	&	$0.74\pm0.23$	&	M7.0 ($2.0-12.5$)	&	M5.0 ($4.0-6.5$)	&	$3010^{+170}_{-230}$	\\
Gl\,852 (B)	&	M4.5 ($3.5-5.5$)	&	M4.0$-$M5.0	&	$0.83\pm0.03$	&	M5.0 ($4.0-5.5$)	&	M4.5 ($4.0-5.5$)	&	$3095^{+85}_{-170}$	\\
Gl\,852 (C)	&	M8.5 ($7.5-9.5$)	&	M7.0$-$M8.0	&	$0.72\pm0.04$	&	M7.5 ($6.5-8.5$)	&	M7.5 ($6.5-8.5$)	&	$2660^{+120}_{-160}$	\\

\hline
\label{SpT}
\end{tabular}
\newline
\end{table*}

\begin{figure}
 \centering
 \includegraphics[width=3.5in]{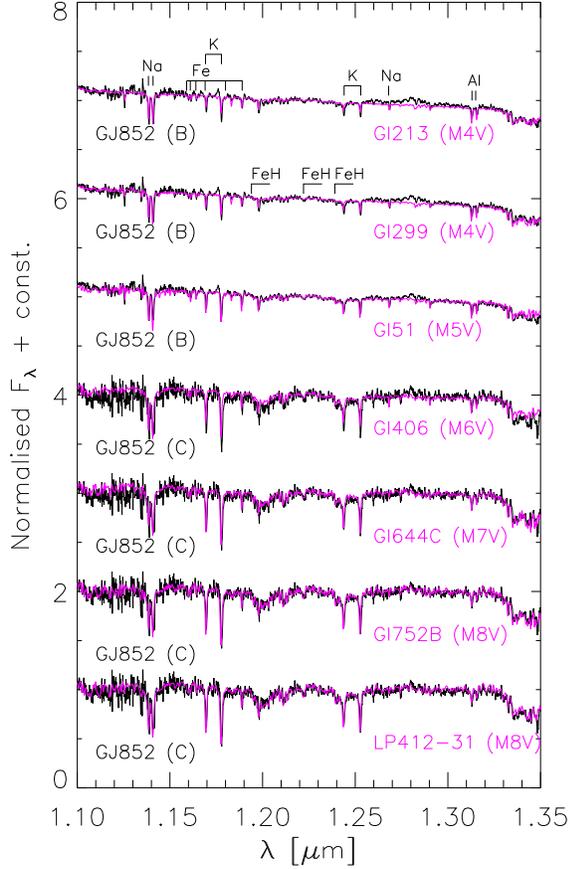} 
    \caption{$J$-band spectra for GJ\,852\,BC. The SINFONI targets are shown in black, and comparison IRTF templates in magenta. }
\label{GJ852_Jbandspectra}
 \end{figure}
 
 \begin{figure}
 \centering
 \includegraphics[width=3.5in]{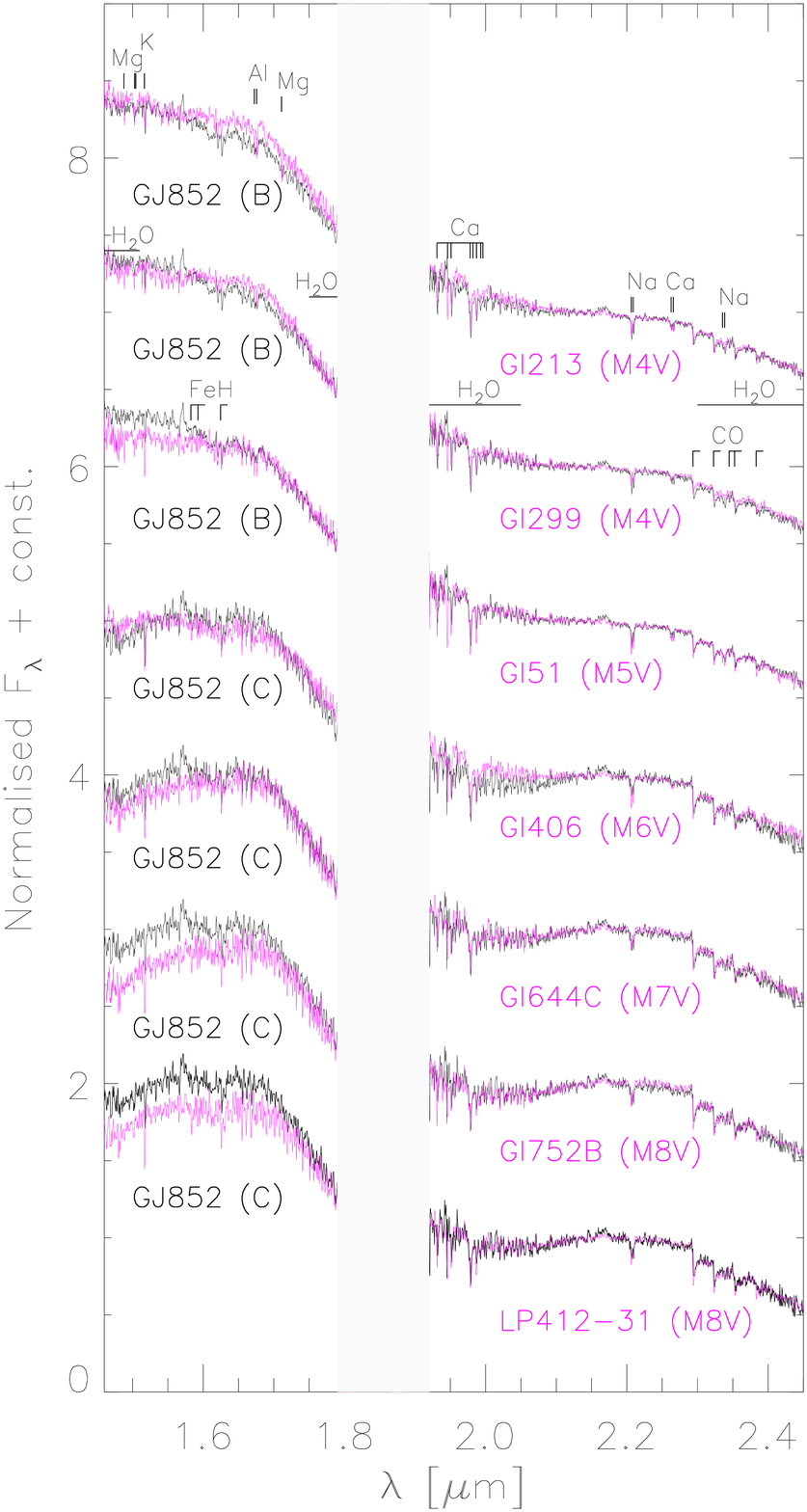} 
    \caption{Similar to Fig. \ref{GJ852_Jbandspectra} but for $H+K$-band.}
\label{GJ852_HKbandspectra}
 \end{figure}
 
 \subsection{Comparison of near-infrared spectral types to AstraLux optical photometry}
In order to derive approximate spectral types for the binary or multiple systems discovered in the AstraLux M dwarf survey and exclude background contaminants, each target was observed in SDSS $i'$- and $z'$-band and individual spectral types were derived from $i'-z'$ colours and the integrated $JHK$ spectral types, with estimated errors of $\pm1$ spectral subtype \citep[see][]{Daemgen2009, Bergfors2010}. 
Obtaining near-infrared spectral types allows us to better quantify the precision of the photometric spectral types, as derived for individual M dwarf binary components in \citet{Bergfors2010, Janson2012, Janson2014a}, and identify potential biases in the method.
We find that, with the exception of J021330\,B, all near-infrared spectral types agree with the photometrically derived spectral types within the photometric $\pm1$ subclass error, and in most cases to within 0.5 subclasses (see Table \ref{SpT}). This confirms that the photometric method used in the AstraLux large M dwarf survey in general provides accurate spectral types.

 \subsection{Comparison with integrated optical spectral types}
 \subsubsection{FEROS observations and data reduction}
 For some of our targets spatially unresolved optical spectroscopy exists, obtained with the Fiberfed Extended Range Optical Spectrograph \citep[FEROS,][]{Kaufer1999} mounted to the ESO-MPG 2.2\,m telescope at La Silla Observatory.
J053018 and J061345 were observed within the ESO programmes 086.A-9014 and 089.A-9013 (Bergfors et al., in preparation), and J024902 within programme 088.A-9032 (Schlieder et al., in preparation). Additional FEROS spectra of J021330 were retrieved from the ESO archive (programme ID 091.C-0216, PI: Rodriguez).  
FEROS provides spectra covering $\rm \lambda=3500-9200\AA$ across 39 orders at $\rm R\approx48000$, using two $2.\arcsec0$ optical fibres separated by $2.\arcmin9$. The targets were observed in `object+sky' mode with one fibre on the star and the other on sky, and at airmass close to 1.0. Observational details are provided in Table \ref{optSpT}. 

The data were reduced using the FEROS Data Reduction System (DRS) within the \begin{small}ESO-MIDAS\end{small} package. The package follows standard spectroscopic reduction procedures which include flat-fielding, background subtraction, bad pixel correction, optimal order extraction, and wavelength calibration using ThAr lamp lines. The software also computes the barycentric velocity correction and re-bins and merges the orders to produce a continuous spectrum. Calibrations used during the reduction were acquired in daytime. For J024902, the RV standard GJ\,1094 was observed on the previous night as an internal calibration check. 

\subsubsection{Optical spectral types}

We determined integrated optical spectral types for these systems by comparing TiO and CaH band head strengths to M dwarf spectral templates from the Sloan Digital Sky Survey \citep{Bochanski2007}. Section 4.1.2 provides details on  additional analysis performed for J024902.

An earlier optical spectral type of $\rm \sim M2$ is found for the M4+M4 system J024902 compared to the photometric resolved optical spectral types, due to the inclusion of the early M primary in the FEROS fibre aperture (J024902\,A, $\rho_{A,BC}\approx0.5\arcsec$, see also Sect. 4.1.2). The optical spectral types are otherwise fully consistent with the AstraLux photometric spectral types. Our optical and near-infrared spectral types agree within errors, which is to be expected on average.

\begin{table*}
\caption{FEROS observational details and optical spectral types.}
\centering
\begin{tabular}{l r r r r}
\hline\hline
ID		&	Obs. Date 			& Optical SpT	&	Exp. Time					\\
		&						&			&	[s]						\\
\hline

J021330 	&	2013-09-23			& 	M4$-$M5	&	4 $\times$ 600				\\
J024902	&	2011-12-06			&	M2		&	1 $\times$ 2300				\\
J053018 	&	2010-11-25, 2012-08-22	&	M3$-$M4	&	1 $\times$ 900, 1 $\times$ 900	\\
J061345	&	2010-11-22, 2012-08-29	&	M4		&	1 $\times$ 900, 1 $\times$ 900	\\

\hline
\label{optSpT}
\end{tabular}
\newline
\end{table*}

\subsection{Surface gravity and chromospheric activity}
Age is one of the most difficult stellar parameters to determine. 
A combination of several properties is usually required to determine youth, since many signs may suggest, however not establish by themselves, that a star is young.
For M dwarfs, such spectral features can include chromospheric and coronal activity (emission features, in particular strong $\rm H\alpha$ emission, X-ray emission, flares), signs of accretion discs (e.g. photometric excess, forbidden O \begin{small}I\end{small} emission lines), Li-absorption in the optical spectrum at 6708\AA. Other signs of youth include low surface gravity, 
low tangential velocity, and kinematic properties consistent with known Young Moving Groups (YMGs) or associations.

\begin{figure}
 \centering
 \includegraphics[width=3.5in]{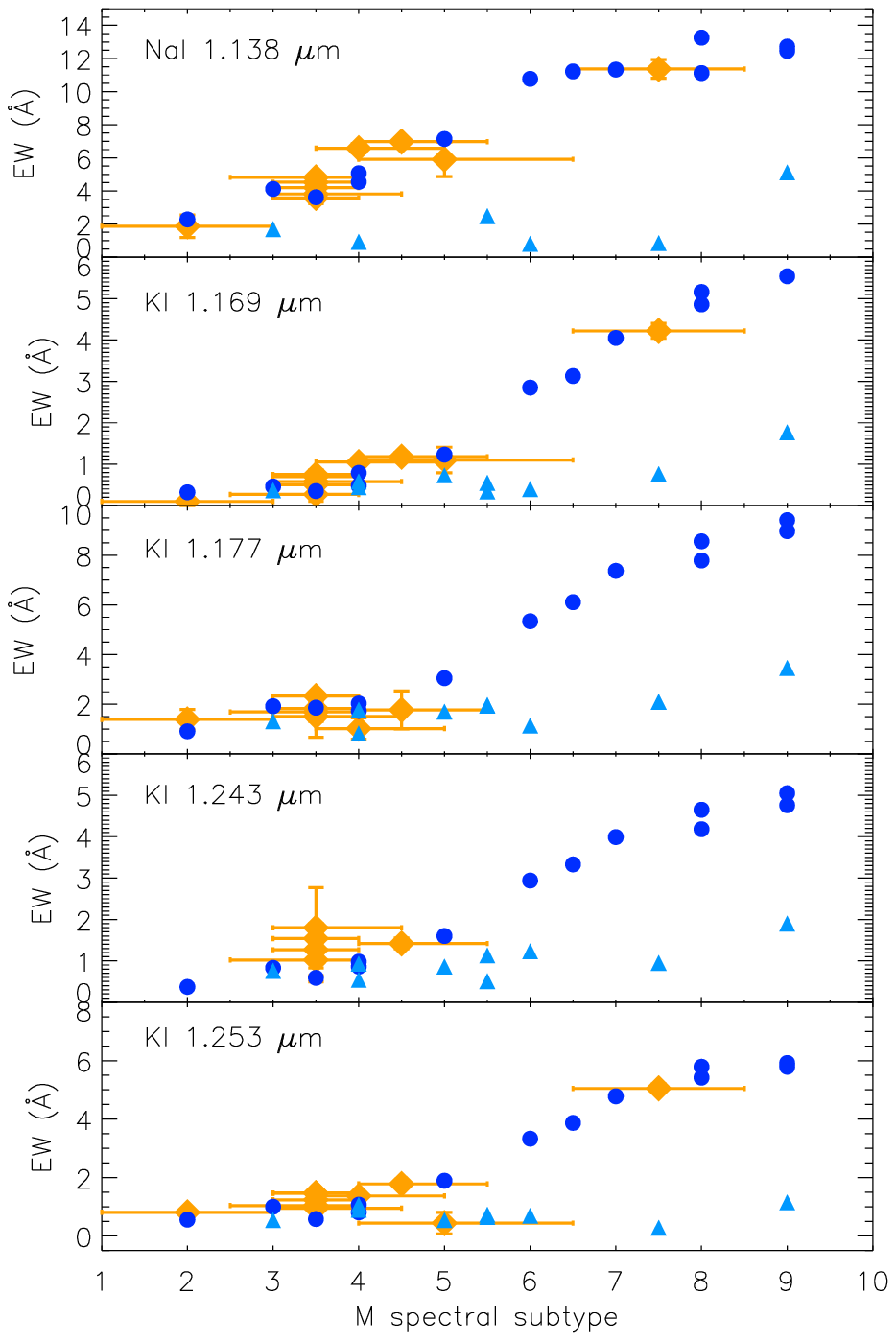} 
    \caption{EWs of surface gravity sensitive features. The SINFONI EWs are shown as yellow diamonds, and compared to EWs measured for IRTF field star templates (dark blue dots) and young stars from the \citet{Manara2013} sample of $<$10\,Myr old stars (light blue triangles).}
\label{EWs}
 \end{figure}

Low surface gravity can be measured in medium resolution spectra such as ours for spectral types M5 or later using gravity sensitive alkali lines in the near-infrared \citep[see e.g.][]{Gorlova2003, McGovern2004, Kirkpatrick2006}, or by using spectral indices \citep[e.g.][]{Lucas2001, Kirkpatrick2006, Allers2007}. We measured the equivalent widths (EW) of the gravity sensitive Na \begin{small}I\end{small} doublet at 1.138$\mu$m and the K \begin{small}I\end{small} doublets at $1.169,1.177\mu$m and $1.243,1.253\mu$m with associated errors as described in \citet{Bonnefoy2014}, with the psuedo-continuum wavelength regions reported therein. For a given spectral subtype, low surface gravity, hence youth, can be seen in the reduced strength of the alkali lines compared to main sequence stars. Table \ref{EW} lists the measured EWs for all stars in our sample, with the exception of J040805 and J024902 for which the quality of the spectra was not sufficient for precise measurements. These are also plotted in Fig. \ref{EWs}, together with EWs for field dwarfs from the IRTF SpeX library and $<$10\,Myr old stars from \citet{Manara2013} for comparison. We measured the EWs for all stars in our sample for consistency, even though the only stars in our target sample that are classified as $\rm \sim M5$ or later are the stars in the 
GJ\,852\,BC system and J061610\,B. For the latter, the measured Na \begin{small}I\end{small} at 1.138$\mu$m is slightly weaker than for the field main sequence template, and the K \begin{small}I\end{small} at 1.169 and 1.253$\mu$m overlap with the young template measurement, suggesting intermediate surface gravity. We see no indication of particularly weak alkali lines in the $J$-band spectra for the other targets, and hence no signs of low surface gravity and youth. 
 
No emission lines indicating youth, such as Br$\gamma$ or Pa$\beta$, can be confidently detected in our near-infrared spectra. Weak `bumps' can be seen in the $J$-band spectra of J061345 and J061610 at roughly the position of Pa$\beta$, however these features likely arise from incomplete H-absorption line subtraction in the telluric spectra during data reduction. Similar `bumps' are seen around the Br$\gamma$ absorption feature in $K$-band for GJ\,852\,BC.

\begin{table*}
\caption{Equivalent widths of gravity sensitive features.}
\centering
\begin{tabular}{l r r r r r}
\hline\hline
ID	&	Na\begin{small}I\end{small} (1.138$\,\mu$m)	&	K\begin{small}I\end{small} (1.169$\,\mu$m)	&	K\begin{small}I\end{small} (1.177$\,\mu$m)	&	K\begin{small}I\end{small} (1.243$\,\mu$m)	&	K\begin{small}I\end{small} (1.253$\,\mu$m) 	\\
	&	(\AA)					&	(\AA)					&	(\AA)					&	(\AA)					&	(\AA)					\\	
\hline
	
J021330 (A)	&	$4.20\pm0.27$	&	$0.70\pm0.06$	&	$2.33\pm0.14$	&	$1.27\pm0.63$	&	$1.46\pm0.11$		\\
J021330 (B)	&	$3.58\pm0.35$	&	$0.50\pm0.22$	&	$1.83\pm0.36$	&	$1.54\pm0.11$	&	$1.47\pm0.10$		\\
J053018 (A)	& 	$1.87\pm0.68$ &	$0.10\pm0.10$ &	$1.39\pm0.40$ &	--			&	$0.81\pm0.06$		\\
J053018 (B)	&	$4.53\pm0.31$ &	$0.75\pm0.11$ &	--			& 	--			&	$1.24\pm0.05$		\\
J061345 (A)	&	$4.82\pm0.23$	&	$0.27\pm0.17$	&	$1.69\pm0.38$	&	$1.02\pm0.53$	&	$1.04\pm0.07$		\\
J061345 (B)	&	$6.58\pm0.28$	&	$1.05\pm0.09$	&	$1.02\pm0.43$	&	--			&	$1.37\pm0.11$		\\
J061610 (A)	&	$3.82\pm0.36$	&	$0.58\pm0.26$	&	$1.51\pm0.84$	&	$1.80\pm0.97$	&	$0.95\pm0.07$		\\
J061610 (B)	&	$5.91\pm1.04$	&	$1.10\pm0.31$	&	--			&	--			&	$0.44\pm0.37$		\\
GJ\,852 (B)	&	$6.98\pm0.24$	&	$1.18\pm0.10$	&	$1.77\pm0.76$	&	$1.42\pm0.13$	&	$1.78\pm0.06$		\\
GJ\,852 (C)	&	$11.37\pm0.56$&	$4.22\pm0.18$	&	--			&	--			&	$5.05\pm0.15$		\\

\hline
\label{EW}
\end{tabular}
\end{table*}

\section{Discussion}
\subsection{Additional youth indicators}
\subsubsection{Candidate members of YMGs}
Based on the selection of targets from their strong coronal emission and low tangential velocity, \citet{Riaz2006} estimated that their catalog, from which the AstraLux target sample was obtained, consisted of mainly stars younger than $\approx$600\,Myr, and using the age-velocity relation of \citet{Holmberg2009} we derived an upper age for our AstraLux sample of $\la$1\,Gyr in \citet{Bergfors2010}. Since then, the kinematics of several of our original large survey targets have been investigated for kinematic membership in YMGs and some are candidate members of e.g. the AB Dor Moving Group (MG), the $\beta$ Pic MG, and the Argus and Columba associations \citep[][Schlieder et al., in preparation]{Malo2013,Malo2014a, Rodriguez2013}, including five of our SINFONI targets: 

\textbf{J021330:}
A convergent point analysis by \citet{Rodriguez2013} yielded a high probability (87\%) of $\beta$ Pic MG membership, however, their analysis using the BANYAN statistical software \citep{Malo2013} placed this target as a field system. The  updated BANYAN  \begin{small}II\end{small}  web tool \citep{Malo2013, Gagne2014}, which assumes a refined prior based on the expected populations, places this system as part of the field population. FEROS spectra obtained from the ESO archive show Balmer line and Ca \begin{small}II\end{small} H \& K emission, but no Li-absorption.

{\bf J024902:} See Sect. 4.1.2.

\textbf{J053018:}
This triple system is flagged by \citet{Malo2014a} as a candidate member of the AB Dor MG, with a probability of 97.7\% when including radial velocity measurements in the analysis. Using the BANYAN  \begin{small}II\end{small}  tool, we find a 73.9\% probability of kinematic membership using the refined prior. Additional integrated FEROS spectra show that the system is chromospherically active with strong Balmer line and Ca \begin{small}II\end{small} H \& K emission, however no visible Li-absorption (Bergfors et al., in preparation).

\textbf{J061345:}
The binary has 99.99\% probability of belonging to the Argus association when radial velocity is included \citep{Malo2014a}, decreasing to 76.1\% with the BANYAN \begin{small}II\end{small} tool. No Li-absorption is visible in integrated FEROS spectra (Bergfors et al., in preparation).

\textbf{J061610:}
Using the BANYAN  \begin{small}II\end{small}  tool with non-uniform priors, we find a 4\% probability that the system belongs to the $\beta$ Pic MG \citep[see also][]{Janson2014b}, while if using uniform prior the probability increases to 99.6\%. No radial velocity measurements were included in either analysis. The convergent point analysis tool of \citet{Rodriguez2013} finds a probability of 84\% of $\beta$ Pic MG membership. 

Our BANYAN  \begin{small}II\end{small}  analysis of the GJ\,852 and J040805 systems places them as field objects. 

\subsubsection{FEROS observations of the J024902 system}

The CASTOFFS survey to identify young, low-mass stars near the Sun included 2MASS J02490228-1029220\,ABC as a candidate of the $\beta$ Pic MG  \citep[][Schlieder et al. in preparation]{Schlieder2012}. The star was selected as a candidate on the basis of consistent position and proper motion, and strong X-ray and UV emission.

{\bf Radial and Rotational Velocity:}
To measure the radial velocity (RV) and rotational velocity of J024902\,ABC, we used \begin{small}IDL\end{small} software to perform a cross-correlation (CC) analysis \citep{Mazeh2002, BenderSimon2008} with a suite of $\rm K5-M5$ RV templates taken from \citet{Prato2002}. These template stars were observed during FEROS runs in 2011 December and 2012 October and were reduced following the same methods as the science target. We measured the RV of J024902\,ABC across portions of four echelle orders chosen to be free of strong telluric absorption. The average RV across the four orders was $\rm 17.1\pm1.1\,km\,s^{-1}$, where the dominant source of error is a $\rm 1\,km\,s^{-1}$ systematic introduced by the use of empirical RV templates. The CC function in each case was strong and single peaked, showing no indication of a tight, spectroscopic binary. The strongest peak was found when using the M2.5 star GJ\,752\,A as a template. We also measured a projected rotational velocity of {\it v} sin {\it i}\,=$\rm 11\pm3\,km\,s^{-1}$ by cross-correlating our spectrum with rotationally broadened templates.

{\bf Spectral type:}
Our RV analysis indicated that J024902\,ABC had a spectral type of $\rm \sim M2$ from the best match RV template. As a further check, we smoothed our spectrum to $\rm R\approx1000$ and performed a visual comparison to the SDSS M dwarf spectral type templates from The Hammer \begin{small}IDL\end{small} spectral typing suite \citep{Covey2007}. Although our spectrum is not flux calibrated, our visual comparison of the strength of atomic and molecular absorption features between 5500 and 7000\AA\,yields a spectral type of $\rm M2\pm1$. We presume this to be the spectral type of the primary, J024902\,A, which contributes more flux than the B and C components.

\begin{figure}
 \centering
 \includegraphics[width=3.5in]{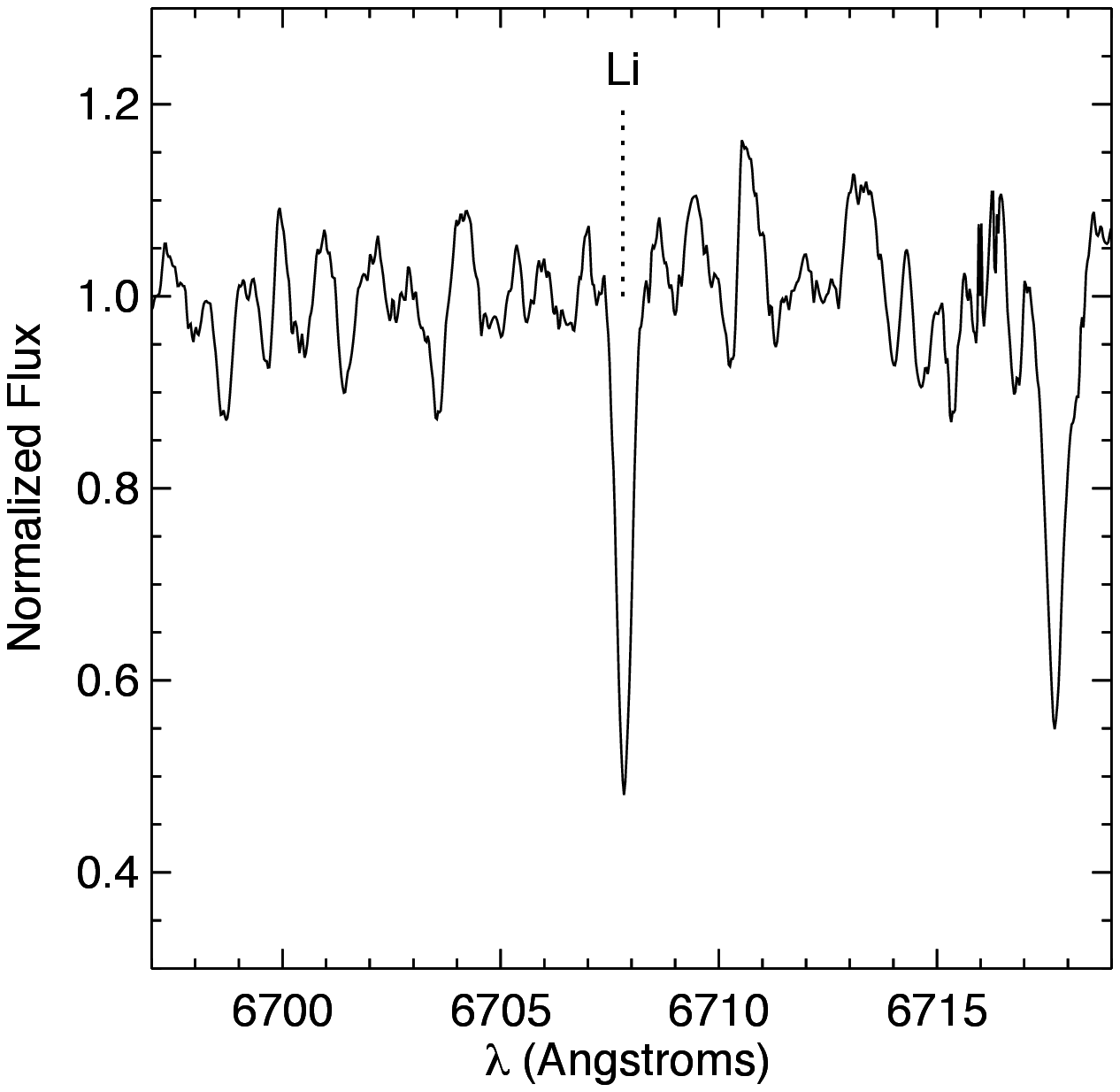} 
    \caption{Li-absorption in FEROS spectrum of J024902\,ABC.}
\label{Li}
 \end{figure}

\begin{figure}
 \centering
 \includegraphics[width=3.5in]{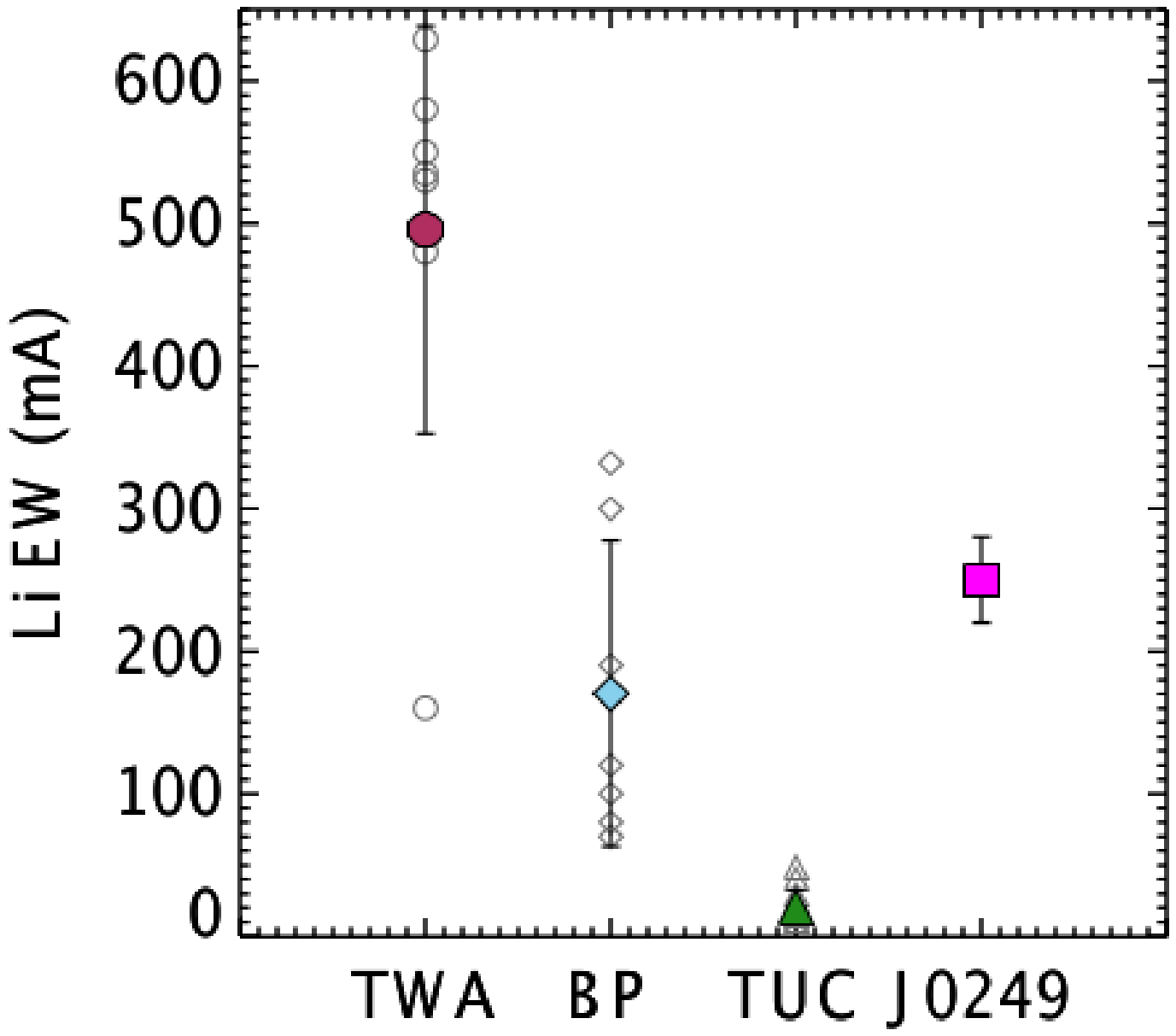} 
    \caption{Li EW for M1$-$M2 members of the TW Hydrae association (TWA), $\beta$ Pic MG (BP), and Tucana-Horologium association (TUC), from \citet{daSilva2009, Kraus2014}. Individual measurements are plotted as gray open symbols while the means and standard deviations are plotted as larger symbols with error bars. The Li EW and error for the J024902 system, in which the $\rm\sim M2$ primary contributes all observed Li, is shown as the magenta square. The measured EW is only consistent with similar spectral type members of the $\beta$ Pic MG.}
\label{Li_comp}
 \end{figure}

{\bf Age indicators and kinematics:}
The FEROS spectrum of J024902\,ABC is rich with emission from the Hydrogen Balmer series ($\rm H\alpha, H\beta, H\gamma, H\delta, H\epsilon$) and other signatures of strong magnetic activity. Additionally, the spectrum exhibits strong Li absorption at $\rm 6707.8\AA$ with $\rm EW=250\pm30$\,m\AA (see Fig. \ref{Li}).
Taking into account the relative fluxes of an M2 and two M4 components at $\rm 6707.8\AA$, the Li strength as a function of temperature, and the Li depletion rate for M2 and M4 dwarfs at ages $\ge20$\,Myr \citep{daSilva2009}, we conclude that the only possible contribution to the Li absorption feature must come from the M2 component. In addition, the Li depletion boundary is in between spectral types M4 and M5 in the $\beta$ Pic MG \citep{BinksJeffries2014, Malo2014b}, and contribution from the lower mass companions is therefore not possible at $\beta$ Pic or older ages. A comparison of our measurement and Li absorption strengths for M1$-$M2 type stars of different ages is shown is Figure \ref{Li_comp}.
The EW of Li in J024902\,A is weaker than in M1$-$M2 type members of the $\approx10$\,Myr TW Hydrae association, but stronger than members of the $\approx40$ Myr old Tucana-Horologium association \citep{Kraus2014}. 
The Li strength is consistent with that of early-M stars in the $\beta$ Pic MG, suggesting a similar age of $\approx20$\,Myr \citep{Mentuch2008, MamajekBell2014}. 

The position and proper motion of the star combined with our RV measurement reveals that the UVW Galactic velocities of the system are consistent with the $\beta$ Pic group distribution for distances between $60-80$\,pc. However, at these distances J024902\,ABC is discrepant with the $\beta$ Pic group XYZ Galactic position distribution by $\approx20$\,pc in both X and Z. This is likely why the Bayesian probability estimator BANYAN  \begin{small}II\end{small} provides negligible probability of membership in $\beta$ Pic or any other MG using the available kinematics. Although the combination of consistent partial kinematics and strong evidence for youth make a compelling case for J024902\,ABC to be a member of the $\beta$ Pic MG, final group membership assignment will require a parallax measurement and a better understanding of the full XYZ distribution of the $\beta$ Pic group. Nevertheless, the unambiguous detection of Li in the optical spectrum of the system indicates an age $<40$\,Myr and the system warrants further study.

\subsection{Hertzsprung-Russell diagram}
Comparison with evolutionary models can provide additional constraints on the system ages and mass estimates to be compared to dynamical masses in the future. We compared the positions of the two systems containing components of spectral type M5 or later, J061610 and GJ\,852\,BC, in a $M_K$ vs. $T_{\rm eff}$ diagram to the \citet{Baraffe2015}, hereafter BHAC15, isochrones.

\begin{figure}
 \centering
 \includegraphics[width=3.5in]{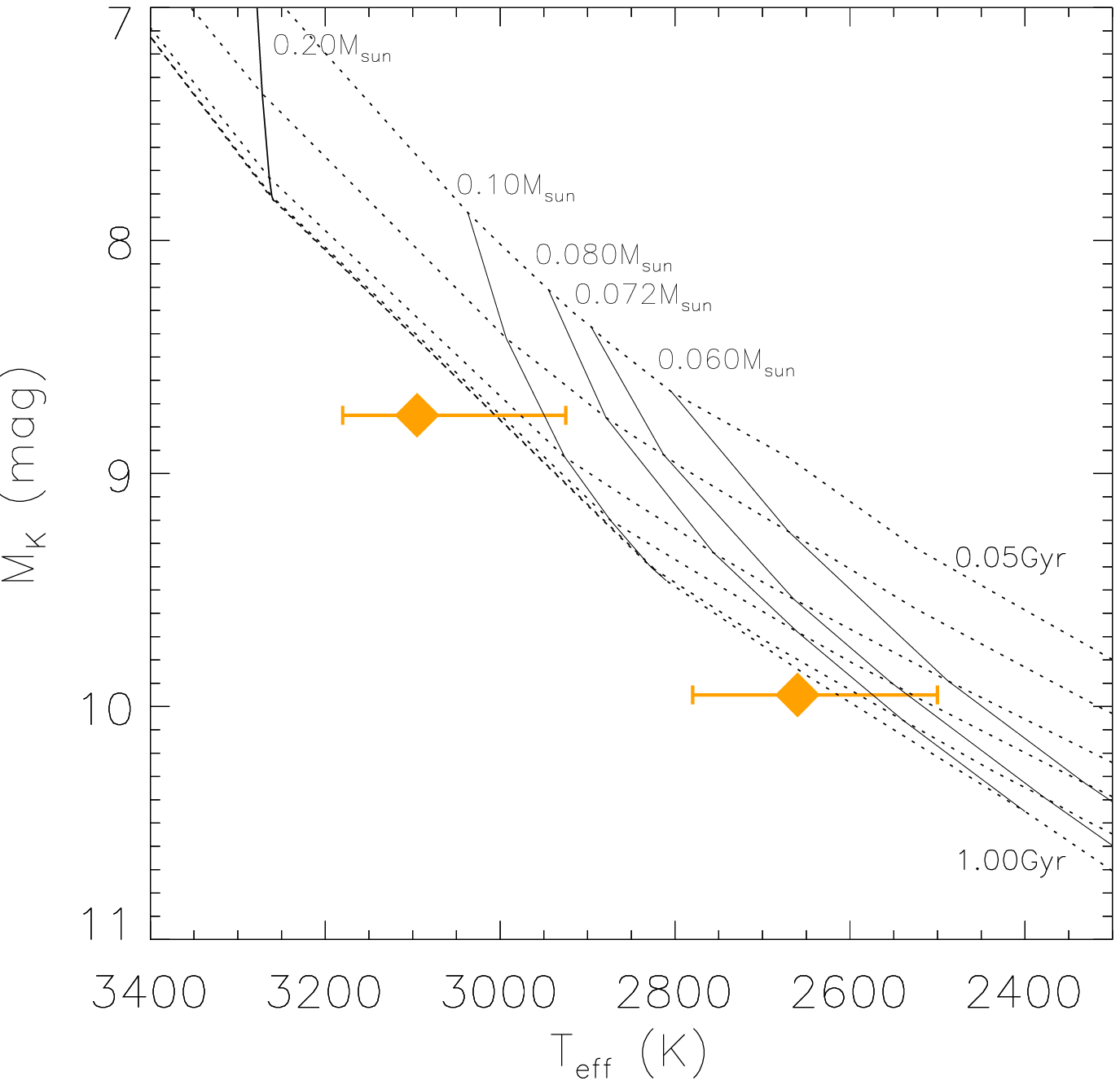} 
    \caption{Comparison of observed properties of the GJ\,852\,BC system to the BHAC15 evolutionary models. Isochrones for ages 0.05, 0.10, 0.20, 0.30, 0.50 and 1.0\,Gyr are shown as dotted lines; evolutionary sequences for constant mass are shown as solid lines.}
\label{HRD}
 \end{figure}

The GJ\,852\,BC system consists of an M4.5 and an M7.5 star and has a trigonometric parallax of $96\pm6$\,mas \citep{HarringtonDahn1980}. This system is of particular interest for orbital monitoring and age determination, since both young and late-type M dwarf binaries with dynamical masses and well characterised spectra are rare.
We measured the $K$-band flux ratio between the BC components and converted to individual apparent magnitudes using integrated 2MASS photometry \citep{Cutri2003}.
The individual component absolute $K$-band magnitudes and derived effective temperatures are plotted together with BHAC15 isochrones and iso-mass contours in Fig. \ref{HRD}. 
 Assuming co-evality, the system is likely older than $\approx$250\,Myr, consistent with the findings from the analysis of gravity sensitive features in the previous section which showed no sign of low surface gravity. For an age $\la300$\,Myr, GJ\,852\,C has a model mass below the hydrogen burning limit ($0.072\,M_{\sun}$).

\begin{figure}
 \centering
 \includegraphics[width=3.5in]{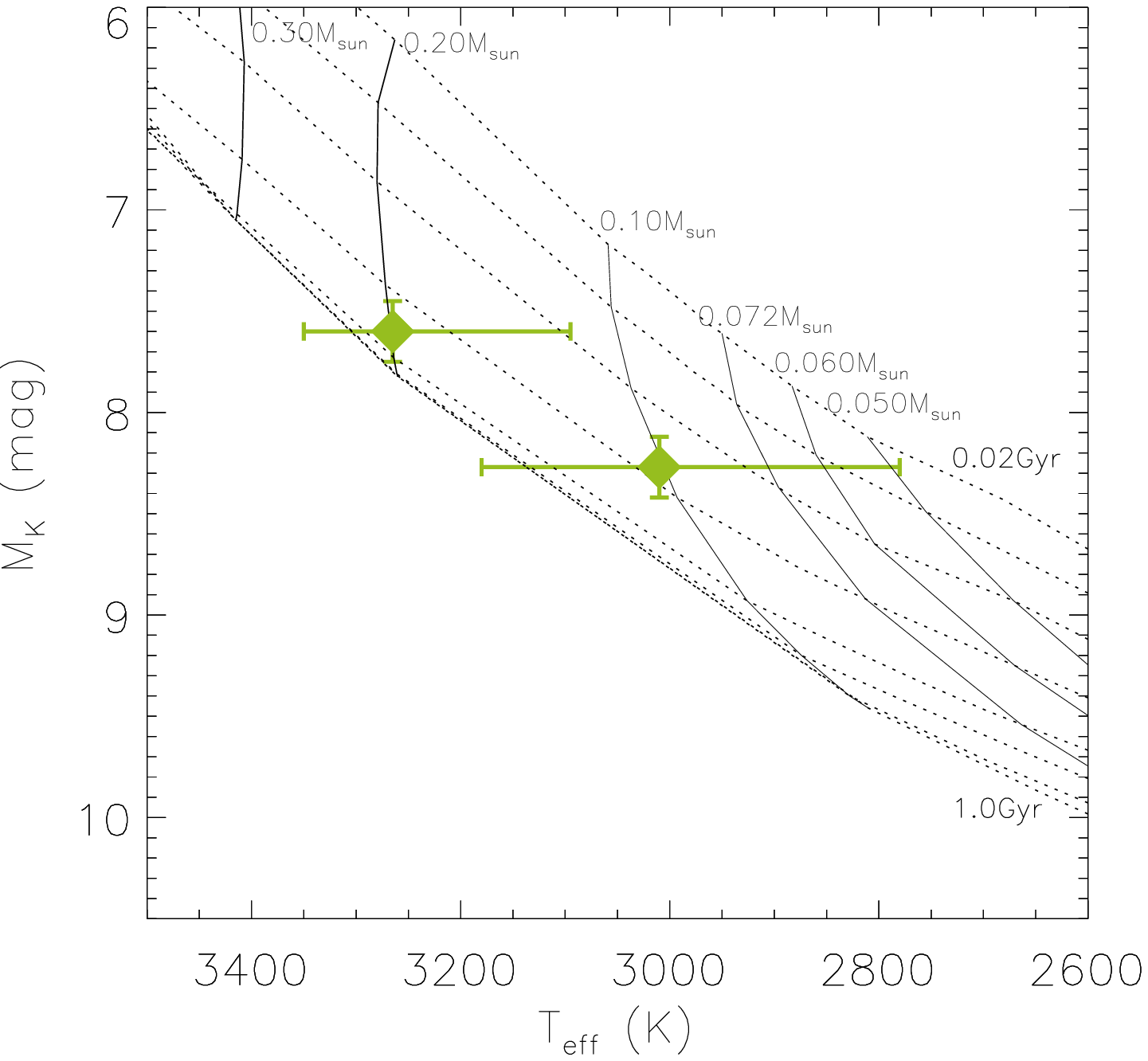} 
    \caption{Similar to Fig. \ref{HRD} but for J061610. Isochrones are shown for ages 0.02, 0.03, 0.05, 0.10, 0.20, 0.30, 0.50 and 1.0\,Gyr.}
\label{HRD_0616}
 \end{figure}

Because of its intermediate surface gravity results in the EW analysis and its debated $\beta$ Pic MG membership \citep{Malo2013, Janson2014b}, we also put the J061610 system on an H-R diagram, despite its lack of a trigonometric parallax measurement. We assumed the spectroscopic (statistical) distance of $47\pm7$\,pc derived by \citet{Malo2013} and plot the derived $M_K$ vs. $T_{\rm eff}$ and BHAC15 isochrones in Figure \ref{HRD_0616}. Assuming co-evality and considering the more stringent constraints set by the primary star parameters, we find a $1\sigma$ system age of $\ga50$\,Myr. At this age, both components in the system have masses above the hydrogen burning limit. Various age estimates for the $\beta$ Pic MG in the literature based on different methods such as the lithium depletion boundary, kinematics and ischronal fitting constrains the age to the range $10-40$\,Myr \citep[see e.g. ][for a summary]{MamajekBell2014}. A system age in this interval can not be rejected from this analysis. We note that for young ages our observed spectral types correspond to higher $T_{\rm eff}$ than we assumed using the \citet{KrausHillenbrand2007} empirical models for Praesepe ($\approx600$\,Myr), shifting the stars to the left in the figure and implying a higher age.

\section{Summary and conclusions}
Our near-infrared spectral analysis of seven close binaries from the AstraLux M dwarf survey provides spectral types for all stars and constraints on surface gravity and age for two systems: GJ\,852\,BC and J061610. The spectral types derived from visual comparison of spectral energy distribution and absorption features and measuring the \citet{Rojas-Ayala2012} $\rm H_2O-K2$ index were found consistent with the estimates determined from $i'-z'$ colours in the AstraLux M dwarf survey, thereby validating the photometric method. Additional optical spectra of J024902 provides radial and rotational velocity and primary star spectral type as well as age constraints.

The ages of these systems are of particular value for future dynamical mass studies.
Upper limits of $\la1$\,Gyr were derived in \citet{Bergfors2010}, and lower limits were here explored for the two systems containing components with spectral type M5 or later from EW measurements of gravity sensitive alkali lines and comparison of observed parameters with theoretical isochrones.

From comparison with the BHAC15 evolutionary models we found a $1\sigma$ lower age limit of $\ga250$\,Myr for GJ\,852. This intermediate age lower limit is consistent with the analysis of gravity sensitive $J$-band alkali lines and the shape of the $H+K$ spectra, neither of which showed any sign of low gravity and hence ongoing contraction.
For an age $\la300$\,Myr we find that the model predicted mass of the C component is below the hydrogen burning limit. 

The J061610 system lacks a trigonometric parallax measurement, however, assuming the spectroscopic distance derived by \citet{Malo2013}, we found a $1\sigma$ system age of $\ga50$\,Myr. An age consistent with the $\beta$\,Pic MG, in which the system is a kinematic member candidate, can not be ruled out from this analysis. The EW analysis suggests  intermediate surface gravity.
Future measurements (e.g. the Li EW, radial velocity and trigonometric parallax) would provide firmer constraints on age and model predicted masses. 

An analysis of optical spectra and kinematics of the J024902 system suggests an age of $\la40$\,Myr, based on the strong Li absorption. Further data such as parallax is needed to establish membership in the $\beta$ Pic MG.

The ultimate goal of this binary study is to better constrain evolutionary models, 
which have been shown to systematically under predict masses for $M\la0.5M_{\sun}$ 
objects \citep{HillenbrandWhite2004}.
The AstraLux Large M Dwarf Survey, from which these binaries were selected for 
spectroscopic characterisation, discovered $\approx200$ nearby, low-mass binary systems, 
many of which belong to YMGs. 
Ongoing orbital monitoring together with 
precise parallactic distances obtained with {\it Gaia} \citep{Perryman2001} 
will within a few years provide dynamical masses for a large number of binaries, providing 
stringent constraints 
on evolutionary models for young, low-mass stars.

\section*{Acknowledgments}
We thank the staff at the Paranal and La Silla observatories for their support, and Jay Farihi and the anonymous referee for useful comments and suggestions.

\label{lastpage}

\end{document}